\documentclass{PoS}
\usepackage{amsmath}
\usepackage{mathtools}
\usepackage{amssymb}
\usepackage{graphicx}
\usepackage{todonotes}

\newcommand{\bear}{\begin{eqnarray}}
\newcommand{\eear}{\end{eqnarray}}

\newcommand{\mL}{\mathcal{L}}

\title{
Collider phenomenology of HEFT\\ and new $Wh$ and $Zh$ resonances
}

\ShortTitle{Collider phenomenology of HEFT}

\author{Antonio Dobado, \speaker{Felipe J. Llanes-Estrada}, and Juan J. Sanz Cillero \thanks{Work supported by Spanish grants MINECO:FPA2011-27853-C02-01,  FPA2014-53375-C2-1-P and FPA2016-75654-C2-1-P, and carried out in the inspiring atmosphere of the theoretical physics department and UPARCOS.}\\
Departamento de Fisica Teorica I, Plaza de las Ciencias 1, Fac. CC. Fisicas; Universidad Complutense de Madrid, 28040, Madrid, Spain.\\
        E-mail: \email{fllanes@fis.ucm.es}}

\abstract{
 We report computations of the production cross-section of longitudinal electroweak and Higgs boson pairs within Effective Field Theory for the Electroweak sector (including the Higgs). 
 We have recently reported theoretical studies of  gauge boson-gauge boson and Higgs-Higgs resonance production with various quantum numbers.
 We are now focusing  on gauge boson-Higgs boson two-body axial-vector resonances and show a typical cross section.
 At last, we point out that  photon-photon production has also been studied (in $e^-e^+$ as well as pp machines), as this is a very clean process allowing access to scalar and tensor resonances. 
}

\FullConference{EPS-HEP 2017, European Physical Society conference on High Energy Physics\\
		5-12 July 2017\\
		Venice, Italy}

\begin{document}
Deviations from the electroweak standard model can be encoded in the Lagrangian of Higgs Effective Field Theory  (HEFT), which in the TeV resonance region (thus, neglecting the masses $m_W\sim m_Z\sim m_h\sim 100$ GeV) can be written as~\cite{Delgado:2015kxa}
\begin{eqnarray} \label{bosonLagrangian}
{\cal L}  =  \frac{1}{2}\left(1 +2 a \frac{h}{v} +b\left(\frac{h}{v}\right)^2\right)
\partial_\mu \omega^a
\partial^\mu \omega^b\left(\delta_{ab}+\frac{\omega^a\omega^b}{v^2}\right)   
\nonumber +\frac{1}{2}\partial_\mu h \partial^\mu h 
  +  \frac{4 a_4}{v^4}(\partial_\mu \omega^a\partial_\nu \omega^a)^2 \nonumber  \\
+\frac{4 a_5}{v^4}( \partial_\mu \omega^a\partial^\mu \omega^a)^2
+\frac{g}{v^4} (\partial_\mu h \partial^\mu h )^2  
  + \frac{2 d}{v^4} \partial_\mu h\partial^\mu h\partial_\nu \omega^a  \partial^\nu\omega^a
+\frac{2 e}{v^4} \partial_\mu h\partial^\nu h\partial^\mu \omega^a \partial_\nu\omega^a.
\end{eqnarray}
In that energy region, the Equivalence Theorem equates longitudinal gauge--boson  $W_L$, \, $Z_L$ scattering amplitudes by the corresponding $\omega, \, z$ Goldstone--boson ones.  In a separate contribution to these proceedings~\cite{Llanes-Estrada:2017ozu} we briefly discussed how the LHC could confront this potential--discovery region if its energy proved insufficient. But should Nature have arranged for new resonances to be light enough (a few TeV), we need theory input to see what LHC cross sections are expected.

Recently~\cite{Dobado:2015hha} we computed the production cross-section of both vector and scalar resonances at both LHC and lepton colliders. The cross sections found were small but reachable if the new resonances were below 2 TeV. These theoretical computations were then extended to full Monte Carlo simulations in~\cite{Delgado:2017cls}.
The conclusion is that to obtain a usable number of events if a new $WZ$ resonance would sit below 2 TeV, or any number of events at all if above, 3000 fb$ ^{-1}$ of luminosity would need to be collected at the LHC in leptonic channels (see table 3 in~\cite{Delgado:2017cls}). 
Thus, there is room for much future improvement in analysis profiting from $W$ and $Z$ jet tagging techniques.

In this contribution we turn to upcoming work on  axial-vector resonances in the range up to 3 TeV; in QCD-like theories, these axial vector states are heavier than the corresponding vector ones because the quantum numbers $J^P=1^-$ can be obtained with a fermion and an antifermion in an s-wave, whereas the $1^+$ quantum numbers require a p-wave.

\begin{figure}
\centering
\raisebox{-0.5\height}{\includegraphics*[width=0.3\textwidth]{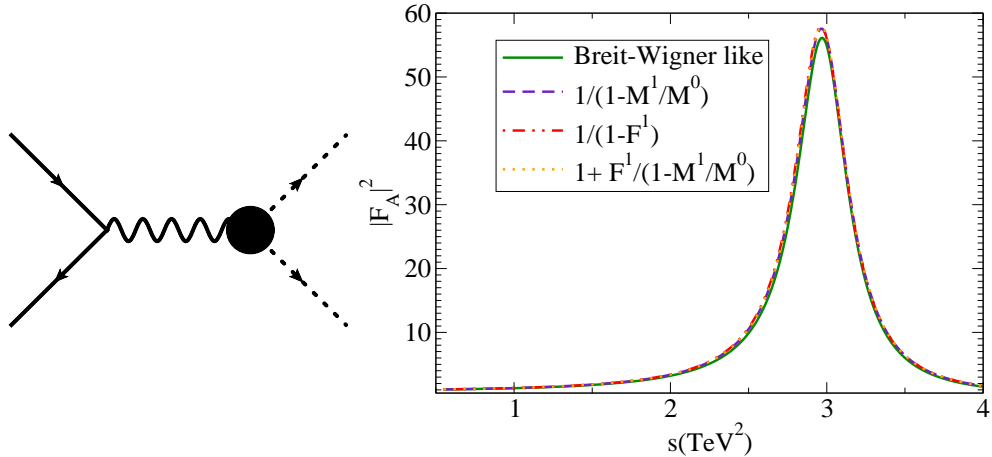}}
\raisebox{-0.5\height}{\includegraphics*[width=0.55\textwidth]{FactoresForma.eps}}
\caption{\label{fig:FFs} {\small {\bf Left:} Production mechanism, annihilating a quark-antiquark pair from the initial $pp$ at the LHC. The resulting transversal $W$ couples to the Electroweak Symmetry Breaking Sector with an axial-vector form factor (black blob) derived from HEFT, and that can be resonant. {\bf Right:} Several models of that axial-vector form factor that feature a resonance with a mass of about 3 TeV.}}
\end{figure}

We have employed several models of the axial vector form factor in $\omega h$ which will be discussed in our upcoming publication~\cite{Cilleroprep} and plotted them in figure~\ref{fig:FFs}. 

\noindent They are constructed from the chiral expansions of the elastic amplitude for $\omega h\to \omega h$, $M\simeq M^{(0)} + M^{(1)}$ and the form factor $F\simeq F^{(0)} + F^{(1)}$ both to one loop. These are then unitarized according to prescriptions well--known from hadron physics. 
The agreement between the different models is impressive as long as the resonance is not too broad, and any of them can be used with confidence.
Knowledge of the form factor allows us to calculate the process $pp \to W_T + X\to W_L +h + X$ where the final state $W_L$ and $h$ may rescatter (as described by the form factor) and resonate.

Once this elementary-particle cross section $\hat{\sigma}(s)$ is at hand, we convolute it with the 
proton parton distribution functions 
(we follow Owens, Accardi and Melnitchouk's with nuclear and finite-$Q^2$ corrections~\cite{Owens:2012bv}) to obtain $pp$ production cross sections at the LHC with an energy $E_{\rm tot}=13$ TeV,
\begin{equation} \label{ppintermediateW}
\frac{d\sigma}{ds}(pp\to W_Lh(s)+X) = \int_\frac{s}{E^2_{\rm tot}} \frac{dx_u}{x_u E^2_{\rm tot}} \hat{\sigma}(s) F_p(x_u)
F_p(x_{\bar{d}}) 
\, , 
\qquad \mbox{with $x_{\bar{d}} = s/(x_uE^2_{\rm tot})$.}
\end{equation}

\begin{figure}
\centering
\includegraphics*[width=0.55\textwidth]{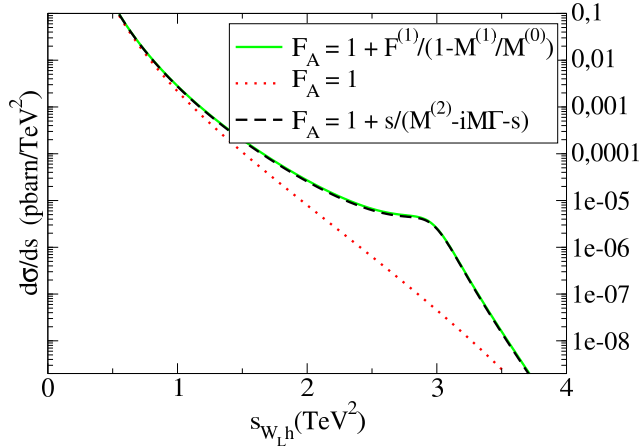}
\caption{\label{fig:Xsecwh} {\small Production cross section of $W_L h\simeq \omega h$ pairs with (top) and without (bottom) a 3 TeV axial-vector resonance. The cross-section is enhanced by the latter, in this case by one and a half orders of magnitude.}}
\end{figure}

\noindent The computation is reported in figure~\ref{fig:Xsecwh}.
There, we inject a resonance of mass 3 TeV and width 0.4 TeV with two of the form factors from figure~\ref{fig:FFs}. This is achieved by the parameters $a=0.95$, $b=0.7a^2$ (away from their SM values $a=b=1$), $e(\mu=3{\rm TeV})=1.64\times 10^{-3}$ and $f_9=-0.6\times 10^{-2}$; this last one appears at NLO upon coupling an axial current to the Lagrangian of Eq.~(\ref{bosonLagrangian}).

The figure shows very clearly the resonance as a shoulder in the differential cross-section, and even if the experimental uncertainty bars do not allow to resolve it, the cross-section is enhanced over the non-resonant Standard Model ($F_A=1$) by at least one order of magnitude or more.

\begin{figure}
\centering
\includegraphics*[width=0.65\textwidth]{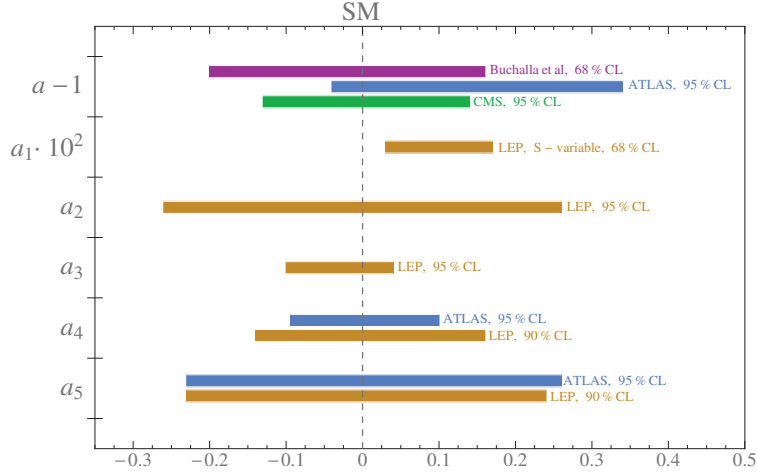}
\caption{\label{fig:parameters}    
  {\small  Existing constraints, from LEP precision studies as well as from existing LHC searches, on the parameters of the HEFT Lagrangian.  
  Figure borrowed from~\cite{Delgado:2017cls} (see references therein).} } 
\end{figure}

Such resonance searches, even if they return empty handed, can be useful to constrain  the parameters of HEFT, by binding the maximum value that their separation from the SM can take.
Meanwhile, we collect in figure~\ref{fig:parameters} a number of constraints from the literature~\cite{Delgado:2017cls} (employing direct searches as well as LEP precision studies of the S,T and U Peskin-Takeuchi parameters).

Finally, we would like to point out that photon-photon initial states can be used as a very clean probe of both scalar and tensor resonances, not directly accessible by the mechanism of figure~\ref{fig:FFs} as the intermediate $W_T$ or $Z_T$ projects over vector or axial-vector quantum numbers. 
The Lagrangian that enhances HEFT by adding $\gamma\gamma$ couplings
\begin{multline}
\mL_4 = %
\frac{e^2a_1}{2v^2}A_{\mu\nu}A^{\mu\nu}\left(v^2 - 4\omega^+\omega^-\right) %
 + \frac{2e(a_2-a_3)}{v^2}A_{\mu\nu}\left[%
         i\left(\partial^\nu\omega^+\partial^\mu\omega^- - \partial^\mu\omega^+\partial^\nu\omega^- \right) %
   \right. \\ \left. %
        +eA^\mu\left( \omega^+\partial^\nu\omega^- + \omega^-\partial^\nu\omega^+ \right)
        -eA^\nu\left( \omega^+\partial^\mu\omega^- + \omega^-\partial^\mu\omega^+ \right)
        \right] -\frac{c_{\gamma}}{2}\frac{h}{v}e^2 A_{\mu\nu} A^{\mu\nu} .
\end{multline}
and the corresponding amplitudes, both in HEFT perturbation theory and unitarized with the Inverse Amplitude and the N/D methods, have been constructed in~\cite{Delgado:2014jda} 
(in that equation, $e$ is the positron electric charge and not the NLO parameter). 
In an upcoming work~\cite{IvanMiguel} we will present cross-section estimates for both $pp$ and $e^-e^+$ processes. Suffice it here to say that the $\gamma\gamma$ induced processes give very small production rates for resonances above 1 TeV, {\it e.g.} a broad resonance at 1.4 TeV generated with $a_4(3{\rm TeV)}=10^{-3}$ yields
$d\sigma(\gamma\gamma\to \omega\omega)/dp_t^2\sim 0.05 $fb/TeV$ ^2$ at $p_t=200$GeV, and this becomes very hard to observe upon convoluting with the photon flux factors in the proton.



\begin{thebibliography}{99}



\bibitem{Delgado:2015kxa}
  R.~L.~Delgado, A.~Dobado and F.~J.~Llanes-Estrada,
  Phys.\ Rev.\ D {\bf 91} (2015) no.7,  075017.

\bibitem{Llanes-Estrada:2017ozu}
  F.~J.~Llanes-Estrada, R.~L.Delgado and A.~Dobado,
  proceedings of EPS-HEP 17, to be published by PoS,
  arXiv:1709.10491 [hep-ph].


\bibitem{Dobado:2015hha}
  A.~Dobado, F.~K.~Guo and F.~J.~Llanes-Estrada,
  Commun.\ Theor.\ Phys.\  {\bf 64} (2015) 701.



\bibitem{Delgado:2017cls}
  R.~L.~Delgado, A.~Dobado, D.~Espriu, C.~Garcia-Garcia, M.~J.~Herrero, X.~Marcano and J.~J.~Sanz-Cillero,
  arXiv:1707.04580 [hep-ph].

\bibitem{Cilleroprep}
J.~J.~Sanz-Cillero, A.~Dobado and F.~J.~Llanes-Estrada, 
work in preparation.

\bibitem{Owens:2012bv}
  J.~F.~Owens, A.~Accardi and W.~Melnitchouk,
  Phys.\ Rev.\ D {\bf 87} (2013) no.9,  094012.

\bibitem{Delgado:2014jda}
  R.~L.~Delgado, A.~Dobado, M.~J.~Herrero and J.~J.~Sanz-Cillero,
  JHEP {\bf 1407} (2014) 149; 
  R.~L.~Delgado, A.~Dobado and F.~J.~Llanes-Estrada,
  Eur.\ Phys.\ J.\ C {\bf 77} (2017) no.4,  205.

\bibitem{IvanMiguel} R.L.Delgado {\it et al.}, \emph{``Collider production of EW resonances through intermediate $\gamma\gamma$''}, in preparation.



\end{thebibliography}
\end{document}